\documentclass[pra,aps,showpacs,epsf,floats]{revtex4}

\usepackage{amssymb}
\usepackage{amsfonts}
\usepackage{amsmath}

\usepackage{graphicx}

\begin{document}

\title{Concatenation of Error Avoiding with Error Correcting Quantum Codes
for Correlated Noise Models}
\author{Carlo Cafaro$^{1}$ and Stefano Mancini$^{2}$}
\affiliation{$^{1,2}$School of Science and Technology, Physics Division, University of
Camerino, I-62032 Camerino, Italy}

\begin{abstract}
We study the performance of simple error correcting and error avoiding
quantum codes together with their concatenation for correlated noise models.
Specifically, we consider two error models: i) a bit-flip (phase-flip) noisy
Markovian memory channel (model I); ii) a memory channel defined as a memory
degree dependent linear combination of memoryless channels with Kraus
decompositions expressed solely in terms of tensor products of $X$-Pauli ($Z$%
-Pauli) operators (model II). The performance of both the three-qubit bit
flip (phase flip) and the error avoiding codes suitable for the considered
error models is quantified in terms of the entanglement fidelity. We
explicitly show that while none of the two codes is effective in the extreme
limit when the other is, the three-qubit bit flip (phase flip) code still
works for high enough correlations in the errors, whereas the error avoiding
code does not work for small correlations. Finally, we consider the
concatenation of such codes for both error models and show that it is
particularly advantageous for model II in the regime of partial correlations.
\end{abstract}

\maketitle

\section{Introduction}

It is well known that one of the most important obstacles in quantum
information processing is decoherence. It causes a quantum computer to lose
its quantum properties destroying its performance advantages over a
classical computer. The unavoidable interaction between the open quantum
processor and its environment corrupts the information stored in the system
and causes errors that may lead to wrong outputs.

There are different methods for preserving quantum coherence. One possible
technique exploits the redundancy in encoding information. This scheme is
known as "quantum error correcting codes" (QECCs). For a comprehensive
introduction to QECCs, we refer to \cite{perimeter}. Within such scheme,
information is encoded in linear subspaces (codes) of the total Hilbert
space in such a way that errors induced by the interaction with the
environment can be detected and corrected. The QECC approach may be
interpreted as an active stabilization of a quantum state in which, by
monitoring the system and conditionally carrying on suitable operations, one
prevents the loss of information. Another possible approach is the so-called
"decoherence free subspaces" (DFSs) (also known as error avoiding or
noiseless codes). For a comprehensive introduction to DFSs, we refer to \cite%
{lidar03}. It turns out that for specific open quantum systems (noise models
in which all qubits can be considered symmetrically coupled with the same
environment), it is possible to design states that are hardly corrupted
rather than states that can be easily corrected (in this sense, DFSs\ are
complementary to QECCs). In other words, it is possible to encode
information in linear subspaces such that the dynamics restricted to such
subspaces is unitary. This implies that no information is loss and quantum
coherence is maintained. DFS is an example of passive stabilization of
quantum information.

The formal mathematical description of the qubit-environment interaction is
usually given in terms of quantum channels. Quantum error correction is
usually developed under the assumption of i.i.d. (identically and
independently distributed) errors. These error models are characterized by
memoryless communication channels $\Lambda $ such that $n$-channel uses is
given by $\Lambda ^{\left( n\right) }=\Lambda ^{\otimes n}$. In such cases
of complete independent decoherence, qubits interact with their own
environments which do not interact with each other. However, in actual
physical situations, qubits do interact with a common environment which
unavoidably introduces correlations in the noise. For instance, there are
situations where qubits in a ion trap set-up are collectively coupled to
their vibrational modes \cite{garg96}. In other situations, different qubits
in a quantum dot design are coupled to the same lattice, thus interacting
with a common thermal bath of phonons \cite{loss98}. The exchange of bosons
between qubits causes spatial and temporal correlations that violate the
condition of error independence \cite{hwang01}. Memory effects introduce
correlations among channel uses with the consequence that $\Lambda ^{\left(
n\right) }\neq \Lambda ^{\otimes n}$. Recent studies try to characterize the
effect of correlations on the performance of QECCs \cite{clemens04,
klesse05, shabani08, d'arrigo08, carlo-PLA}. It appears that correlations
may have negative \cite{klesse05} or positive \cite{shabani08} impact on
QECCs depending on the features of the error model being considered.

Devising new good quantum codes for both independent and correlated noise
models is a highly non trivial problem. However, it is usually possible to
manipulate existing codes to construct new ones suitable for more general
error models and with higher performances \cite{gotty96, calder97}.
Concatenation is perhaps one of the most successful quantum coding tricks
employed to produce new codes from old ones. The concept of concatenated
codes was first introduced in classical error correcting schemes by Forney 
\cite{forney66}. Roughly speaking, concatenation is a method of combining
two codes (an inner and an outer code) to form a larger code. In classical
error correction, Forney has carried on extensive studies of concatenated
codes, how to choose the inner and outer codes, and what error probability
threshold values can be achieved. The first applications of concatenated
codes in quantum error correction appear in \cite{gotty97, knill96}. In the
quantum setting, concatenated codes play a key role in fault tolerant
quantum computation and in constructing good degenerate quantum error
correcting codes. For instance, Shor's degenerate code can be constructed by
concatenating the three-qubit bit flip and phase flip codes.

It is known that DFSs are efficient under conditions in which each qubit
couples to the same environment (collective decoherence) while ordinary
QECCs are designed to be efficient when each individual qubit couples to a
different environment (independent decoherence) \cite{lidar99}. While none
of the two error correction schemes is effective in the extreme limit when
the other is, QECCs will still work for correlated errors \cite{gotty96-1,
knill97}, whereas DFSs will not work in the independent error case \cite%
{lidar98}. Therefore, it would be of interest to study the performance of
concatenated codes obtained by combining DFSs and QECCs for explicit noise
models in the presence of partially correlated errors. Indeed, this will be
one of the main purposes in our work.

In this article, we study the performance of simple error correcting and
error avoiding quantum codes together with their concatenation for
correlated noise models. Specifically, we consider two error models: i) a
bit-flip (phase-flip) noisy Markovian memory channel (model I); ii) a memory
channel defined as a memory degree dependent linear combination of
memoryless channels with Kraus decompositions expressed solely in terms of
tensor products of $X$-Pauli ($Z$-Pauli) operators (model II). The
performance of both the three-qubit bit flip (phase flip) and the error
avoiding codes suitable for the considered error models is quantified in
terms of the entanglement fidelity. We explicitly show that while none of
the two codes is effective in the extreme limit when the other is, the
three-qubit bit flip (phase flip) code still works for high enough
correlations in the errors, whereas the error avoiding code does not work
for small correlations. Finally, we consider the concatenation of such codes
for both error models and show that it is particularly advantageous for
model II in the regime of partial correlations.

The layout of this article is as follows. In Section II, we briefly discuss
about the concatenation technique in quantum coding, the DFSs and, the
entanglement fidelity as a performance measure of quantum error correction
schemes. In Section III, we evaluate the performances of the three-qubit bit
flip code and suitable DFSs for error models I\ and II. The performance of
the concatenated code for both models appears in Section IV. Finally, in
Section V, we present our concluding remarks.

\section{On Concatenation, Decoherence Free Subspaces and, Entanglement
Fidelity}

In this Section, we briefly discuss about the concatenation technique in
quantum coding, the DFSs and, the entanglement fidelity.

\subsection{Quantum Concatenation Trick}

For the sake of simplicity we only use two layers of concatenation and
consider single qubit encoding. The generalization to arbitrary $l$ layers
of concatenation and multi-qubit encoding is relatively straightforward \cite%
{gaitan}. Assume the inner code (first layer) is a $\left[ n_{1}\text{, }%
k_{1}\text{, }d_{1}\right] $ stabilizer code $\mathcal{C}_{1}$ with
generators $G_{1}=\left\{ g_{i}^{1}:i=1\text{,..., }n_{1}-k_{1}\right\} $,
and the outer code (second layer) is a $\left[ n_{2}\text{, }1\text{, }d_{2}%
\right] $ stabilizer code $\mathcal{C}_{2}$ with generators $G_{2}=\left\{
g_{j}^{2}:j=1\text{,..., }n_{j}-1\right\} $. The concatenated code $\mathcal{%
C}\overset{\text{def}}{=}\mathcal{C}_{1}\circ \mathcal{C}_{2}$ is to map $%
k_{1}$ qubits into $n=n_{1}n_{2}$ qubits, with code construction parsing the 
$n$ qubits into $n_{1}$ blocks $B\left( b\right) $ ($b=1$,..., $n_{1}$) each
containing $n_{2}$ qubits. In other words, given a codeword $\left\vert c_{%
\text{inner}}\right\rangle $ for the inner code $\mathcal{C}_{1}$,%
\begin{equation}
\left\vert c_{\text{inner}}\right\rangle =\sum_{j=0}^{k_{1}-1}\alpha
_{j}\left\vert \phi _{j}\right\rangle \text{,}
\end{equation}%
where $\left\{ \left\vert \phi _{j}\right\rangle \right\} $ are basis
vectors for $\mathcal{C}_{1}$, the concatenated code $\mathcal{C}$ is
constructed as follows. For any codeword $\left\vert c_{\text{outer}%
}\right\rangle $ for the outer code $\mathcal{C}_{2}$,%
\begin{equation}
\left\vert c_{\text{outer}}\right\rangle =\sum_{i_{1}\text{,..., }%
i_{n_{2}}}\alpha _{i_{1}\text{,..., }i_{n_{2}}}\left\vert i_{1}\text{...}%
i_{n_{2}}\right\rangle \text{,}
\end{equation}%
with $\left\vert i_{1}\text{...}i_{n_{2}}\right\rangle =\left\vert
i_{1}\right\rangle \otimes $...$\otimes \left\vert i_{n_{2}}\right\rangle $,
replace each basis vector $\left\vert i_{j}\right\rangle $ with $i_{j}=0$%
,..., $k_{1}-1$ for $j=1$,..., $n_{2}$ by a basis vector $\left\vert \phi
_{i_{j}}\right\rangle $ in $\mathcal{C}_{1}$, that is%
\begin{equation}
\left\vert c_{\text{concatenated}}\right\rangle \overset{\text{def}}{=}%
\sum_{i_{1}\text{,..., }i_{n_{2}}}\alpha _{i_{1}\text{,..., }%
i_{n_{2}}}\left\vert \phi _{i_{1}}\right\rangle \otimes \text{...}\otimes
\left\vert \phi _{i_{n_{2}}}\right\rangle \text{.}
\end{equation}%
Further details on the construction of the stabilizer generators of $%
\mathcal{C}$ can be found in \cite{gaitan}. As a final remark, we simply
point out that the above mentioned construction produces a $\left[ n_{1}n_{2}%
\text{, }k_{1}\text{, }d\right] $ code with $d\geq d_{1}d_{2}$. As an
illustrative example, consider a concatenated code that is based on $l$
layers of concatenation of the same code, for instance the seven-qubit $%
\left[ 7\text{, }1\text{, }3\right] $ CSS\ code. Here, an unencoded qubit is
encoded into a block of seven qubits. Next each qubit is itself encoded into
a block of seven qubits. Repeating this process $l$ times produces the
concatenated code in which one logical qubit is recursively encoded into the
state of $7^{l}$ physical qubits. Thus, such concatenated code is a $\left[
7^{l}\text{, }1\text{, }d\right] $ code with $d\geq 3^{l}$. Recall that a
distance $d$ QECC can correct $t$ errors, where $t=\left[ \frac{\left(
d-1\right) }{2}\right] $ and $\left[ x\right] $ is the integer part of $x$.
Since $d\geq 3^{l}$, we see that the number of errors that a concatenated
code can correct grows exponentially with the number of layers of
concatenation $l$. Therefore, concatenation can be a good coding trick as
long as the error model at each encoding level has the same form, that is,
the same Kraus error operators with possibly different amplitude.

\subsection{Decoherence Free Subspaces}

Following \cite{lidar03}, we mention few relevant properties of DFSs.
Consider the dynamics of a closed system composed of a quantum system $%
\mathcal{Q}$ coupled to a bath $\mathcal{B}$. The unitary evolution of the
closed system is described by the combined system-bath Hamiltonian $H_{\text{%
tot}}$,%
\begin{equation}
H_{\text{tot}}=H_{\mathcal{Q}}\otimes I_{\mathcal{B}}+H_{\mathcal{B}}\otimes
I_{\mathcal{Q}}+H_{\text{int}}\text{, }H_{\text{int}}=\sum_{\alpha
}E_{\alpha }\otimes B_{\alpha }\text{.}  \label{def}
\end{equation}%
The operator $H_{\mathcal{Q}}$ ($H_{\mathcal{B}}$) is the system (bath)
Hamiltonian, $I_{\mathcal{Q}}$ ($I_{\mathcal{B}}$) is the identity operator
of the system (bath), $E_{\alpha }$ are the error generators acting solely
on $\mathcal{Q}$ while $B_{\alpha }$ act on the bath. The last term in (\ref%
{def}) is the interaction Hamiltonian.

A subspace $\mathcal{H}_{\text{DFS}}$ of the total system Hilbert space $%
\mathcal{H}$ is a decoherence free subspace (unitary evolution in $\mathcal{H%
}_{\text{DFS}}$ for all possible bath states) if and only if:

\begin{enumerate}
\item $E_{\alpha }\left\vert \psi \right\rangle =c_{\alpha }\left\vert \psi
\right\rangle $ with $c_{\alpha }\in 
%TCIMACRO{\U{2102} }%
%BeginExpansion
\mathbb{C}
%EndExpansion
$, for all states $\left\vert \psi \right\rangle $ spanning $\mathcal{H}_{%
\text{DFS}}$, and for every error operator $E_{\alpha }$ in $H_{\text{int}}$%
. In other words, all basis states spanning $\mathcal{H}_{\text{DFS}}$ are
degenerate eigenstates of all the error generators $E_{\alpha }$;

\item $\mathcal{Q}$ and $\mathcal{B}$ are initially decoupled;

\item $H_{\mathcal{Q}}\left\vert \psi \right\rangle $ has no overlap with
states in the subspace orthogonal to $\mathcal{H}_{\text{DFS}}$.
\end{enumerate}

To establish a direct link between QECCs and DFSs, it is more convenient to
present an alternative formulation of DFSs in terms of the Kraus operator
sum representation. Within such description, the evolution of the system $%
\mathcal{Q}$ density matrix is written as,%
\begin{equation}
\rho _{\mathcal{Q}}\left( t\right) =Tr_{\mathcal{B}}\left[ U\left( \rho _{%
\mathcal{Q}}\otimes \rho _{\mathcal{B}}\right) U^{\dagger }\right]
=\sum_{k}A_{k}\rho _{\mathcal{Q}}\left( 0\right) A_{k}^{\dagger }\text{,}
\end{equation}%
where $U=e^{-\frac{i}{\hbar }H_{\text{tot}}t}$ is the unitary evolution
operator for the system-bath closed system and the Kraus operator $A_{k}$
(satisfying the normalization condition) are given by,%
\begin{equation}
A_{k}=\sqrt{n}\left\langle m|U|n\right\rangle \text{ with }%
\sum_{k}A_{k}^{\dagger }A_{k}=I_{\mathcal{Q}}\text{,}
\end{equation}%
where $k=\left( n\text{, }m\right) $, $\left\vert m\right\rangle $ and $%
\left\vert n\right\rangle $ are bath states. It turns out that a $N_{\text{%
DFS}}$-dimensional subspace $\mathcal{H}_{\text{DFS}}$ of $\mathcal{H}$ is a
DFS if and only if all Kraus operators have an identical unitary
representation (in the basis where the first $N_{\text{DFS}}$ states span $%
\mathcal{H}_{\text{DFS}}$) upon restriction to it, up to a multiplicative
constant,%
\begin{equation}
A_{k}=\left( 
\begin{array}{cc}
g_{k}U_{\mathcal{Q}}^{\left( \text{DFS}\right) } & 0 \\ 
0 & \bar{A}_{k}%
\end{array}%
\right) \text{,}  \label{3}
\end{equation}%
where $g_{k}=\sqrt{n}\left\langle m|U_{\text{c}}|n\right\rangle $ and $U_{%
\text{c}}=e^{-\frac{i}{\hbar }H_{\text{c}}t}$ with $H_{\text{c}}=H_{\mathcal{%
B}}+H_{\text{int}}$. Furthermore, $\bar{A}_{k}$ is an arbitrary matrix that
acts on $\mathcal{H}_{\text{DFS}}^{\perp }$ (with $\mathcal{H}=\mathcal{H}_{%
\text{DFS}}\oplus \mathcal{H}_{\text{DFS}}^{\perp }$) and may cause
decoherence there; $U_{\mathcal{Q}}^{\left( \text{DFS}\right) }$ is $U_{%
\mathcal{Q}}$ restricted to $\mathcal{H}_{\text{DFS}}$. Now recall that in
ordinary QECCs, it is possible to correct the errors induced by a given set
of Kraus operators $\left\{ A_{k}\right\} $ if and only if,%
\begin{equation}
R_{r}A_{k}=\left( 
\begin{array}{cc}
\lambda _{rk}I_{\mathcal{C}} & 0 \\ 
0 & B_{rk}%
\end{array}%
\right) \text{, }\forall \text{ }r\text{ and }k\text{,}  \label{1}
\end{equation}%
or, equivalently,%
\begin{equation}
A_{k}^{\dagger }A_{k^{\prime }}=\left( 
\begin{array}{cc}
\gamma _{kk^{\prime }}I_{\mathcal{C}} & 0 \\ 
0 & \bar{A}_{k}^{\dagger }\bar{A}_{k^{\prime }}%
\end{array}%
\right) \text{,}  \label{2}
\end{equation}%
where $\left\{ R_{r}\right\} $ are the recovery operators. The first block
in the RHS of (\ref{1}) acts on the code space $\mathcal{C}$ while the
matrices $B_{rk}$ act on $\mathcal{C}^{\perp }$ where $\mathcal{H}=$ $%
\mathcal{C}\oplus \mathcal{C}^{\perp }$. From (\ref{3}) and (\ref{1}), it
follows that DFS can be viewed as a special class of QECCs, where upon
restriction to the code space $\mathcal{C}$, all recovery operators $R_{r}$
are proportional to the inverse of the system $\mathcal{Q}$ evolution
operator,%
\begin{equation}
R_{r}\propto \left( U_{\mathcal{Q}}^{\left( \text{DFS}\right) }\right)
^{\dagger }\text{.}  \label{r}
\end{equation}%
Assuming that (\ref{r}) holds, from (\ref{3}) and (\ref{2}) it also turns
out that,%
\begin{equation}
A_{k}\propto U_{\mathcal{Q}}^{\left( \text{DFS}\right) }\text{,}
\end{equation}%
upon restriction to $\mathcal{C}$. Furthermore, from (\ref{3}) and (\ref{2}%
), it follows that $\gamma _{kk^{\prime }}=g_{k}^{\ast }g_{k^{\prime }}$.
However, while in the QECCs case $\gamma _{kk^{\prime }}$ is in general a
full-rank matrix (non-degenerate code), in the DFSs case this matrix has
rank $1$. In conclusion, a DFS can be viewed as a special type of QECC,
namely a completely degenerate quantum error correcting code where upon
restriction to the code subspace all recovery operators are proportional to
the inverse of the system evolution operator. As a side remark, in view of
this last observation we point out that it is not unreasonable to quantify
the performance of both active and passive QEC schemes by means of the same
performance measure.

\subsection{Entanglement Fidelity}

We recall the concept of entanglement fidelity as a useful performance
measure of the efficiency of quantum error correcting codes. Entanglement
fidelity is a quantity that keeps track of how well the state and
entanglement of a subsystem of a larger system are stored, without requiring
the knowledge of the complete state or dynamics of the larger system. More
precisely, the entanglement fidelity is defined for a mixed state $\rho
=\sum_{i}p_{i}\rho _{i}=$tr$_{\mathcal{H}_{R}}\left\vert \psi \right\rangle
\left\langle \psi \right\vert $ in terms of a purification $\left\vert \psi
\right\rangle \in \mathcal{H}\otimes \mathcal{H}_{R}$ to a reference system $%
\mathcal{H}_{R}$. The purification $\left\vert \psi \right\rangle $ encodes
all of the information in $\rho $. Entanglement fidelity is a measure of how
well the channel $\Lambda $ preserves the entanglement of the state $%
\mathcal{H}$ with its reference system $\mathcal{H}_{R}$. The entanglement
fidelity is defined as follows \cite{schumacher96},%
\begin{equation}
\mathcal{F}\left( \rho \text{, }\Lambda \right) \overset{\text{def}}{=}%
\left\langle \psi |\left( \Lambda \otimes I_{\mathcal{H}_{R}}\right) \left(
\left\vert \psi \right\rangle \left\langle \psi \right\vert \right) |\psi
\right\rangle \text{,}
\end{equation}%
where $\left\vert \psi \right\rangle $ is any purification of $\rho $, $I_{%
\mathcal{H}_{R}}$ is the identity map on $\mathcal{M}\left( \mathcal{H}%
_{R}\right) $ and $\Lambda \otimes I_{\mathcal{H}_{R}}$ is the evolution
operator extended to the space $\mathcal{H}\otimes \mathcal{H}_{R}$, space
on which $\rho $ has been purified. If the quantum operation $\Lambda $ is
written in terms of its Kraus operator elements $\left\{ A_{k}\right\} $ as, 
$\Lambda \left( \rho \right) =\sum_{k}A_{k}\rho A_{k}^{\dagger }$, then it
can be shown that \cite{nielsen96}, 
\begin{equation}
\mathcal{F}\left( \rho \text{, }\Lambda \right) =\sum_{k}\text{tr}\left(
A_{k}\rho \right) \text{tr}\left( A_{k}^{\dagger }\rho \right)
=\sum_{k}\left\vert \text{tr}\left( \rho A_{k}\right) \right\vert ^{2}\text{.%
}
\end{equation}%
This expression for the entanglement fidelity is very useful for explicit
calculations. Finally, assuming that%
\begin{equation}
\Lambda :\mathcal{M}\left( \mathcal{H}\right) \ni \rho \longmapsto \Lambda
\left( \rho \right) =\sum_{k}A_{k}\rho A_{k}^{\dagger }\in \mathcal{M}\left( 
\mathcal{H}\right) \text{, dim}_{%
%TCIMACRO{\U{2102} }%
%BeginExpansion
\mathbb{C}
%EndExpansion
}\mathcal{H=}N  \label{pla1}
\end{equation}%
and choosing a purification described by a maximally entangled unit vector $%
\left\vert \psi \right\rangle \in \mathcal{H}\otimes \mathcal{H}$ for the
mixed state $\rho =\frac{1}{\text{dim}_{%
%TCIMACRO{\U{2102} }%
%BeginExpansion
\mathbb{C}
%EndExpansion
}\mathcal{H}}I_{\mathcal{H}}$ , we obtain%
\begin{equation}
\mathcal{F}\left( \frac{1}{N}I_{\mathcal{H}}\text{, }\Lambda \right) =\frac{1%
}{N^{2}}\sum_{k}\left\vert \text{tr}A_{k}\right\vert ^{2}\text{.}
\label{nfi}
\end{equation}%
The expression\ in (\ref{nfi}) represents the entanglement fidelity when no
error correction is performed on the noisy channel $\Lambda $ in (\ref{pla1}%
).

\section{Three-qubit bit flip code and DFS}

In this Section, we consider two correlated error models. Error correction
is performed by means of the three-qubit bit flip code and a suitable
decoherence free subspace. The code performance measure used is the
entanglement fidelity.

\emph{Model I}. The first model is a bit flip noisy quantum Markovian memory
channel $\Lambda ^{(n)}(\rho )$ (model I). In explicit terms, we consider $n$
qubits and Markovian correlated errors in a bit flip quantum channel,%
\begin{equation}
\Lambda ^{(n)}(\rho )\overset{\text{def}}{=}\sum_{i_{1}\text{,..., }%
i_{n}=0}^{1}p_{i_{n}|i_{n-1}}p_{i_{n-1}|i_{n-2}}\text{...}%
p_{i_{2}|i_{1}}p_{i_{1}}\left( A_{i_{n}}\otimes \text{...}\otimes
A_{i_{1}}\right) \rho \left( A_{i_{n}}\otimes \text{...}\otimes
A_{i_{1}}\right) ^{\dagger }\text{,}  \label{n-general}
\end{equation}%
where $A_{0}\overset{\text{def}}{=}I$, $A_{1}\overset{\text{def}}{=}X$ are
Pauli operators. Furthermore the conditional probabilities $p_{i_{k}|i_{j}}$
are given by, 
\begin{equation}
p_{i_{k}|i_{j}}=(1-\mu )p_{i_{k}}+\mu \delta _{i_{k}\text{, }i_{j}}\text{,}%
\quad p_{i_{k}=0}=1-p\text{,}\;p_{i_{k}=1}=p\text{,}  \label{aaa1}
\end{equation}%
with,%
\begin{equation}
\sum_{i_{1}\text{,..., }i_{n}=0}^{1}p_{i_{n}|i_{n-1}}p_{i_{n-1}|i_{n-2}}%
\text{...}p_{i_{2|i_{1}}}p_{i_{1}}=1\text{.}
\end{equation}%
To simplify our notation, we may choose to omit the symbol of tensor product
"$\otimes $" in the future, $A_{i_{n}}\otimes $...$\otimes A_{i_{1}}\equiv $ 
$A_{i_{n}}$...$A_{i_{1}}$. Furthermore, we may choose to omit the bar "$\mid 
$" in $p_{i_{k}|i_{j}}$ and simply write the conditional probabilities as $%
p_{i_{k}i_{j}}$.

\emph{Model II}. The second quantum communication channel $\Lambda _{\mu
}^{\left( n\right) }\left( \rho \right) $ (model II) that we consider is a
memory quantum channel defined in terms of a linear combination of simple
memoryless channels with Kraus decompositions expressed in terms of bit-flip
error operators. The coefficients of such combination are dependent on the
memory parameter $\mu $. In explicit terms, we consider the following
channel $\Lambda _{\mu }^{\left( n\right) }\left( \rho \right) $,%
\begin{equation}
\Lambda _{\mu }^{\left( n\right) }\left( \rho \right) \overset{\text{def}}{=}%
a_{0}\left( \mu \right) \Lambda _{0}^{\left( n\right) }\left( \rho \right)
+a_{1}\left( \mu \right) \Lambda _{1}^{\left( n\right) }\left( \rho \right) 
\text{,}  \label{gigi}
\end{equation}%
where $a_{0}\left( \mu \right) \overset{\text{def}}{=}1-\mu $, $a_{1}\left(
\mu \right) \overset{\text{def}}{=}\mu $, $\Lambda _{0}^{\left( n\right)
}\left( \rho \right) \equiv \Lambda ^{\left( n\right) }\left( \rho \right) $
in (\ref{n-general}) in the limiting case of $\mu =0$ ($n$-uses of a
memoryless bit flip channel). Finally, $\Lambda _{1}^{\left( n\right)
}\left( \rho \right) $ describes $n$-uses of a memoryless channel whose
Kraus decomposition is characterized only by weight-$0$ and weight-$n$ error
operators with amplitudes $\left( 1-p\right) ^{\frac{n}{2}}$ and $\left[
1-\left( 1-p\right) ^{n}\right] ^{\frac{1}{2}}$, respectively, 
\begin{equation}
\Lambda _{1}^{\left( n\right) }\left( \rho \right) \overset{\text{def}}{=}%
\left( 1-p\right) ^{n}\rho +\left[ 1-\left( 1-p\right) ^{n}\right] X^{1}X^{2}%
\text{...}X^{n-1}X^{n}\rho X^{1}X^{2}\text{...}X^{n-1}X^{n}\text{.}
\end{equation}%
In this Section, QEC is performed via the three-qubit bit flip code and a
suitable decoherence free subspace. Although the error models considered are
not truly quantum in nature, from this preliminary work we hope to gain
useful insights for extending error correction techniques to quantum error
models in the presence of partial correlations. The performance of quantum
error correcting codes is quantified by means of the entanglement fidelity
as function of the error probability $p$ and degree of memory $\mu $.

\subsection{The three-qubit bit flip code}

\subsubsection{Model I}

\emph{Error Operators}. In the simplest example, we consider the limiting
case of (\ref{n-general}) with $n=3$,%
\begin{equation}
\Lambda ^{(3)}(\rho )\overset{\text{def}}{=}\sum_{i_{1}\text{, }i_{2}\text{, 
}i_{3}=0}^{1}p_{i_{3}|i_{2}}p_{i_{2}|i_{1}}p_{i_{1}}\left[
A_{i_{3}}A_{i_{2}}A_{i_{1}}\rho A_{i_{1}}^{\dag }A_{i_{2}}^{\dag
}A_{i_{3}}^{\dag }\right] \text{, with}\sum_{i_{1}\text{, }i_{2}\text{, }%
i_{3}=0}^{1}p_{i_{3}|i_{2}}p_{i_{2}|i_{1}}p_{i_{1}}=1\text{.}  \label{bit}
\end{equation}%
Substituting (\ref{aaa1}) in (\ref{bit}), it follows that the error
superoperator $\mathcal{A}$ associated to channel (\ref{bit}) is defined in
terms of the following error operators,%
\begin{equation}
\mathcal{A}\longleftrightarrow \left\{ A_{0}^{\prime }\text{,.., }%
A_{7}^{\prime }\right\} \text{ with }\Lambda ^{(3)}(\rho )\overset{\text{def}%
}{=}\sum\limits_{k=0}^{7}A_{k}^{\prime }\rho A_{k}^{\prime \dagger }\text{
and, }\sum\limits_{k=0}^{7}A_{k}^{\prime \dagger }A_{k}^{\prime
}=I_{8\times 8}\text{.}  \label{nota}
\end{equation}%
In an explicit way, the error operators $\left\{ A_{0}^{\prime }\text{,.., }%
A_{7}^{\prime }\right\} $ are given by,%
\begin{eqnarray}
A_{0}^{\prime } &=&\sqrt{\tilde{p}_{0}^{\left( 3\right) }}I^{1}\otimes
I^{2}\otimes I^{3}\text{, }A_{1}^{\prime }=\sqrt{\tilde{p}_{1}^{\left(
3\right) }}X^{1}\otimes I^{2}\otimes I^{3}\text{, }A_{2}^{\prime }=\sqrt{%
\tilde{p}_{2}^{\left( 3\right) }}I^{1}\otimes X^{2}\otimes I^{3}\text{, } 
\notag \\
&&  \notag \\
A_{3}^{\prime } &=&\sqrt{\tilde{p}_{3}^{\left( 3\right) }}I^{1}\otimes
I^{2}\otimes X^{3}\text{, }A_{4}^{\prime }=\sqrt{\tilde{p}_{4}^{\left(
3\right) }}X^{1}\otimes X^{2}\otimes I^{3}\text{, }A_{5}^{\prime }=\sqrt{%
\tilde{p}_{5}^{\left( 3\right) }}X^{1}\otimes I^{2}\otimes X^{3}\text{,} 
\notag \\
&&  \notag \\
\text{ }A_{6}^{\prime } &=&\sqrt{\tilde{p}_{6}^{\left( 3\right) }}%
I^{1}\otimes X^{2}\otimes X^{3}\text{, }A_{7}^{\prime }=\sqrt{\tilde{p}%
_{7}^{\left( 3\right) }}X^{1}\otimes X^{2}\otimes X^{3}\text{,}
\end{eqnarray}%
where the coefficients $\tilde{p}_{k}^{\left( 3\right) }$ for $k=1$,.., $7$
are given by,%
\begin{eqnarray}
\tilde{p}_{0}^{\left( 3\right) } &=&p_{00}^{2}p_{0}\text{, }\tilde{p}%
_{1}^{\left( 3\right) }=p_{00}p_{10}p_{0}\text{, }\tilde{p}_{2}^{\left(
3\right) }=p_{01}p_{10}p_{0}\text{, }\tilde{p}_{3}^{\left( 3\right)
}=p_{00}p_{01}p_{1}\text{,}  \notag \\
&&  \notag \\
\text{ }\tilde{p}_{4}^{\left( 3\right) } &=&p_{10}p_{11}p_{0}\text{, }\tilde{%
p}_{5}^{\left( 3\right) }=p_{01}p_{10}p_{1}\text{, }\tilde{p}_{6}^{\left(
3\right) }=p_{01}p_{11}p_{1}\text{, }\tilde{p}_{7}^{\left( 3\right)
}=p_{11}^{2}p_{1}\text{, }  \label{usa1}
\end{eqnarray}%
with,%
\begin{eqnarray}
p_{0} &=&\left( 1-p\right) \text{, }p_{1}=p\text{, }p_{00}=\left( \left(
1-\mu \right) \left( 1-p\right) +\mu \right) \text{, }  \notag \\
&&  \notag \\
p_{01} &=&\left( 1-\mu \right) \left( 1-p\right) \text{, }p_{10}=\left(
1-\mu \right) p\text{, }p_{11}=\left( \left( 1-\mu \right) p+\mu \right) 
\text{.}  \label{usa2}
\end{eqnarray}

\emph{Encoding Operator}. Consider a three-qubit bit flip code that encodes $%
1$ logical qubit into $3$-physical qubits. The codewords are given by,%
\begin{equation}
\left\vert 0\right\rangle \overset{\text{tensoring}}{\longrightarrow }%
\left\vert 0\right\rangle \otimes \left\vert 00\right\rangle =\left\vert
000\right\rangle \overset{\text{def}}{=}\left\vert 0_{\text{L}}\right\rangle 
\text{, }\left\vert 1\right\rangle \overset{\text{tensoring}}{%
\longrightarrow }\left\vert 1\right\rangle \otimes \left\vert
00\right\rangle =\left\vert 100\right\rangle \overset{U_{\text{CNOT}%
}^{12}\otimes I^{3}}{\longrightarrow }\left\vert 110\right\rangle \overset{%
U_{\text{CNOT}}^{13}\otimes I^{2}}{\longrightarrow }\left\vert
111\right\rangle \overset{\text{def}}{=}\left\vert 1_{\text{L}}\right\rangle 
\text{.}  \label{placs}
\end{equation}%
The operator $U_{\text{CNOT}}^{ij}$ is the CNOT\ gate from qubit $i$ to $j$
defined as,%
\begin{equation}
U_{\text{CNOT}}^{ij}\overset{\text{def}}{=}\frac{1}{2}\left[ \left(
I^{i}+Z^{i}\right) \otimes I^{j}+\left( I^{i}-Z^{i}\right) \otimes X^{j}%
\right] \text{.}  \label{you}
\end{equation}%
Finally, the encoding operator $U_{\text{enc}}$ such that $U_{\text{enc}%
}\left\vert 000\right\rangle =\left\vert 000\right\rangle $ and $U_{\text{enc%
}}\left\vert 100\right\rangle =\left\vert 111\right\rangle $ is defined as,%
\begin{equation}
U_{\text{enc}}\overset{\text{def}}{=}\left( U_{\text{CNOT}}^{13}\otimes
I^{2}\right) \circ \left( U_{\text{CNOT}}^{12}\otimes I^{3}\right) \text{.}
\label{bitencoding}
\end{equation}

\emph{Correctable Errors and} \emph{Recovery Operators}. The set of error
operators satisfying the detectability condition \cite{knill02}, $P_{%
\mathcal{C}}A_{k}^{\prime }P_{\mathcal{C}}=\lambda _{A_{k}^{\prime }}P_{%
\mathcal{C}}$, where $P_{\mathcal{C}}=\left\vert 0_{L}\right\rangle
\left\langle 0_{L}\right\vert +$ $\left\vert 1_{L}\right\rangle \left\langle
1_{L}\right\vert $ is the projector operator on the code subspace $\mathcal{C%
}=Span\left\{ \left\vert 0_{L}\right\rangle \text{, }\left\vert
1_{L}\right\rangle \right\} $ is given by,%
\begin{equation}
\mathcal{A}_{\text{detectable}}=\left\{ A_{0}^{\prime }\text{, }%
A_{1}^{\prime }\text{, }A_{2}^{\prime }\text{, }A_{3}^{\prime }\text{, }%
A_{4}^{\prime }\text{, }A_{5}^{\prime }\text{, }A_{6}^{\prime }\right\}
\subseteq \mathcal{A}\text{.}
\end{equation}%
The only non-detectable error is $A_{7}^{\prime }$. Furthermore, since all
the detectable errors are invertible, the set of correctable errors is such
that $\mathcal{A}_{\text{correctable}}^{\dagger }\mathcal{A}_{\text{%
correctable}}$ is detectable. It follows then that,%
\begin{equation}
\mathcal{A}_{\text{correctable}}=\left\{ A_{0}^{\prime }\text{, }%
A_{1}^{\prime }\text{, }A_{2}^{\prime }\text{, }A_{3}^{\prime }\right\}
\subseteq \mathcal{A}_{\text{detectable}}\subseteq \mathcal{A}\text{.}
\end{equation}%
The action of the correctable error operators $\mathcal{A}_{\text{correctable%
}}$ on the codewords $\left\vert 0_{L}\right\rangle $ and $\left\vert
1_{L}\right\rangle $ is given by,%
\begin{eqnarray}
\left\vert 0_{L}\right\rangle &\rightarrow &A_{0}^{\prime }\left\vert
0_{L}\right\rangle =\sqrt{\tilde{p}_{0}^{\left( 3\right) }}\left\vert
000\right\rangle \text{, }A_{1}^{\prime }\left\vert 0_{L}\right\rangle =%
\sqrt{\tilde{p}_{1}^{\left( 3\right) }}\left\vert 100\right\rangle \text{, }%
A_{2}^{\prime }\left\vert 0_{L}\right\rangle =\sqrt{\tilde{p}_{2}^{\left(
3\right) }}\left\vert 010\right\rangle \text{, }A_{3}^{\prime }\left\vert
0_{L}\right\rangle =\sqrt{\tilde{p}_{3}^{\left( 3\right) }}\left\vert
001\right\rangle \text{ }  \notag \\
&&  \notag \\
\left\vert 1_{L}\right\rangle &\rightarrow &A_{0}^{\prime }\left\vert
1_{L}\right\rangle =\sqrt{\tilde{p}_{0}^{\left( 3\right) }}\left\vert
111\right\rangle \text{, }A_{1}^{\prime }\left\vert 1_{L}\right\rangle =%
\sqrt{\tilde{p}_{1}^{\left( 3\right) }}\left\vert 011\right\rangle \text{, }%
A_{2}^{\prime }\left\vert 1_{L}\right\rangle =\sqrt{\tilde{p}_{2}^{\left(
3\right) }}\left\vert 101\right\rangle \text{, }A_{3}^{\prime }\left\vert
1_{L}\right\rangle =\sqrt{\tilde{p}_{3}^{\left( 3\right) }}\left\vert
110\right\rangle \text{.}  \label{ea}
\end{eqnarray}%
The two \ four-dimensional orthogonal subspaces $\mathcal{V}^{0_{L}}$ and $%
\mathcal{V}^{1_{L}}$ of $\mathcal{H}_{2}^{3}$ generated by the action of $%
\mathcal{A}_{\text{correctable}}$ on $\left\vert 0_{L}\right\rangle $ and $%
\left\vert 1_{L}\right\rangle $ are given by,%
\begin{equation}
\mathcal{V}^{0_{L}}=Span\left\{ \left\vert v_{1}^{0_{L}}\right\rangle
=\left\vert 000\right\rangle \text{,}\left\vert v_{2}^{0_{L}}\right\rangle
=\left\vert 100\right\rangle \text{, }\left\vert v_{3}^{0_{L}}\right\rangle
=\left\vert 010\right\rangle \text{, }\left\vert v_{4}^{0_{L}}\right\rangle
=\left\vert 001\right\rangle \text{ }\right\} \text{,}  \label{span1}
\end{equation}%
and,%
\begin{equation}
\mathcal{V}^{1_{L}}=Span\left\{ \left\vert v_{1}^{1_{L}}\right\rangle
=\left\vert 111\right\rangle \text{,}\left\vert v_{2}^{1_{L}}\right\rangle
=\left\vert 011\right\rangle \text{, }\left\vert v_{3}^{1_{L}}\right\rangle
=\left\vert 101\right\rangle \text{, }\left\vert v_{4}^{1_{L}}\right\rangle
=\left\vert 110\right\rangle \right\} \text{,}  \label{span2}
\end{equation}%
respectively. Notice that $\mathcal{V}^{0_{L}}\oplus \mathcal{V}^{1_{L}}=%
\mathcal{H}_{2}^{3}$. The recovery superoperator $\mathcal{R}\leftrightarrow
\left\{ R_{l}\right\} $ with $l=1$,..,$4$ is defined as \cite{knill97},%
\begin{equation}
R_{l}\overset{\text{def}}{=}V_{l}\sum_{i=0}^{1}\left\vert
v_{l}^{i_{L}}\right\rangle \left\langle v_{l}^{i_{L}}\right\vert \text{,}
\label{recovery}
\end{equation}%
where the unitary operator $V_{l}$ is such that $V_{l}\left\vert
v_{l}^{i_{L}}\right\rangle =\left\vert i_{L}\right\rangle $ for $i\in
\left\{ 0\text{, }1\right\} $. Substituting (\ref{span1}) and (\ref{span2})
into (\ref{recovery}), it follows that the four recovery operators $\left\{
R_{1}\text{, }R_{2}\text{, }R_{3}\text{, }R_{4}\right\} $ are given by,%
\begin{eqnarray}
R_{1} &=&\left\vert 0_{L}\right\rangle \left\langle 0_{L}\right\vert
+\left\vert 1_{L}\right\rangle \left\langle 1_{L}\right\vert \text{, }%
R_{2}=\left\vert 0_{L}\right\rangle \left\langle 100\right\vert +\left\vert
1_{L}\right\rangle \left\langle 011\right\vert \text{,}  \notag \\
&&  \notag \\
\text{ }R_{3} &=&\left\vert 0_{L}\right\rangle \left\langle 010\right\vert
+\left\vert 1_{L}\right\rangle \left\langle 101\right\vert \text{, }%
R_{4}=\left\vert 0_{L}\right\rangle \left\langle 001\right\vert +\left\vert
1_{L}\right\rangle \left\langle 110\right\vert \text{.}  \label{rec}
\end{eqnarray}%
Using simple algebra, it turns out that the $8\times 8$ matrix
representation $\left[ R_{l}\right] $ with $l=1$,..,$4$ of the recovery
operators is given by,%
\begin{equation}
\left[ R_{1}\right] =E_{11}+E_{88}\text{, }\left[ R_{2}\right] =E_{12}+E_{87}%
\text{, }\left[ R_{3}\right] =E_{13}+E_{86}\text{, }\left[ R_{4}\right]
=E_{14}+E_{85}\text{, }
\end{equation}%
where $E_{ij}$ is the $8\times 8$ matrix where the only non-vanishing
element is the one located in the $ij$-position and it equals $1$. It
follows that $\mathcal{R}\leftrightarrow \left\{ R_{l}\right\} $ is indeed a
trace preserving quantum operation since,%
\begin{equation}
\sum_{l=1}^{4}R_{l}^{\dagger }R_{l}=I_{8\times 8}\text{.}
\end{equation}%
The action of this recovery operation $\mathcal{R}$ on the map $\Lambda
^{\left( 3\right) }\left( \rho \right) $ in (\ref{nota}) leads to,%
\begin{equation}
\Lambda _{\text{recover}}^{\left( 3\right) }\left( \rho \right) \equiv
\left( \mathcal{R\circ }\Lambda ^{(3)}\right) \left( \rho \right) \overset{%
\text{def}}{=}\sum_{k=0}^{7}\sum\limits_{l=1}^{4}\left( R_{l}A_{k}^{\prime
}\right) \rho \left( R_{l}A_{k}^{\prime }\right) ^{\dagger }\text{.}
\label{pla2}
\end{equation}

\emph{Entanglement Fidelity}. We want to describe the action of $\mathcal{%
R\circ }\Lambda ^{(3)}$ restricted to the code subspace $\mathcal{C}$.
Therefore, we compute the $2\times 2$ matrix representation $\left[
R_{l}A_{k}^{\prime }\right] _{|\mathcal{C}}$ of each $R_{l}A_{k}^{\prime }$
with $l=1$,.., $4$ and $k=0$,.., $7$ where,%
\begin{equation}
\left[ R_{l}A_{k}^{\prime }\right] _{|\mathcal{C}}\overset{\text{def}}{=}%
\left( 
\begin{array}{cc}
\left\langle 0_{L}|R_{l}A_{k}^{\prime }|0_{L}\right\rangle & \left\langle
0_{L}|R_{l}A_{k}^{\prime }|1_{L}\right\rangle \\ 
\left\langle 1_{L}|R_{l}A_{k}^{\prime }|0_{L}\right\rangle & \left\langle
1_{L}|R_{l}A_{k}^{\prime }|1_{L}\right\rangle%
\end{array}%
\right) \text{.}  \label{sopra1}
\end{equation}%
Substituting (\ref{ea}) and (\ref{rec}) into (\ref{sopra1}), it turns out
that the only matrices $\left[ R_{l}A_{k}^{\prime }\right] _{|\mathcal{C}}$
with non-vanishing trace are given by,%
\begin{eqnarray}
\left[ R_{1}A_{0}^{\prime }\right] _{|\mathcal{C}} &=&\sqrt{\tilde{p}%
_{0}^{\left( 3\right) }}\left( 
\begin{array}{cc}
1 & 0 \\ 
0 & 1%
\end{array}%
\right) \text{, }\left[ R_{2}A_{1}^{\prime }\right] _{|\mathcal{C}}=\sqrt{%
\tilde{p}_{1}^{\left( 3\right) }}\left( 
\begin{array}{cc}
1 & 0 \\ 
0 & 1%
\end{array}%
\right) \text{,}  \notag \\
&&  \notag \\
\text{ }\left[ R_{3}A_{2}^{\prime }\right] _{|\mathcal{C}} &=&\sqrt{\tilde{p}%
_{2}^{\left( 3\right) }}\left( 
\begin{array}{cc}
1 & 0 \\ 
0 & 1%
\end{array}%
\right) \text{, }\left[ R_{4}A_{3}^{\prime }\right] _{|\mathcal{C}}=\sqrt{%
\tilde{p}_{3}^{\left( 3\right) }}\left( 
\begin{array}{cc}
1 & 0 \\ 
0 & 1%
\end{array}%
\right) \text{.}
\end{eqnarray}%
Therefore, the entanglement fidelity $\mathcal{F}_{\text{bit}}^{\left(
3\right) }\left( \mu \text{, }p\right) $ defined as,%
\begin{equation}
\mathcal{F}_{\text{bit}}^{\left( 3\right) }\left( \mu \text{, }p\right) 
\overset{\text{def}}{=}\mathcal{F}^{\left( 3\right) }\left( \frac{1}{2}%
I_{2\times 2}\text{, }\mathcal{R\circ }\Lambda ^{(3)}\right) =\frac{1}{%
\left( 2\right) ^{2}}\sum_{k=0}^{7}\sum\limits_{l=1}^{4}\left\vert \text{tr}%
\left( \left[ R_{l}A_{k}^{\prime }\right] _{|\mathcal{C}}\right) \right\vert
^{2}\text{,}  \label{fidel}
\end{equation}%
results,%
\begin{equation}
\mathcal{F}_{\text{bit}}^{\left( 3\right) }\left( \mu \text{, }p\right) =%
\tilde{p}_{0}^{\left( 3\right) }+\tilde{p}_{1}^{\left( 3\right) }+\tilde{p}%
_{2}^{\left( 3\right) }+\tilde{p}_{3}^{\left( 3\right) }\text{.}
\label{usa3}
\end{equation}%
The expression for $\mathcal{F}_{\text{bit}}^{\left( 3\right) }\left( \mu 
\text{, }p\right) $ in (\ref{fidel}) represents the entanglement fidelity
quantifying the performance of the error correction scheme provided by the
three-qubit bit flip code here considered. The quantum operation $\mathcal{%
R\circ }\Lambda ^{(3)}$ appearing in (\ref{fidel}) is defined in equation (%
\ref{pla2}) and the recovery operators $R_{l}$ are explicitly given in (\ref%
{rec}). The action of\textbf{\ }$R_{l}A_{k}^{\prime }$ in (\ref{fidel}) is
restricted to the code space\textbf{\ }$\mathcal{C}$ defined in (\ref{placs}%
).

Substituting (\ref{usa1}) and (\ref{usa2}) into (\ref{usa3}), we finally
obtain%
\begin{equation}
\mathcal{F}_{\text{bit}}^{\left( 3\right) }\left( \mu \text{, }p\right) =\mu
^{2}\left( 2p^{3}-3p^{2}+p\right) +\mu \left( -4p^{3}+6p^{2}-2p\right)
+\left( 2p^{3}-3p^{2}+1\right) \text{ (model I).}  \label{bit1}
\end{equation}%
Notice that for a vanishing degree of memory $\mu $, the entanglement
fidelity becomes,%
\begin{equation}
\mathcal{F}_{\text{bit}}^{\left( 3\right) }\left( 0\text{, }p\right)
=2p^{3}-3p^{2}+1\text{.}
\end{equation}

\emph{Remarks on the coding for phase flip memory channels}. The code for
the phase flip channel has the same characteristics as the code for the bit
flip channel. These two channels are unitarily equivalent since there is a
unitary operator, the Hadamard gate $H$, such that the action of one channel
is the same as the other, provided the first channel is preceded by $H$ and
followed by $H^{\dagger }$ \cite{nielsen00},%
\begin{equation}
\Lambda ^{\text{phase}}\left( \rho \right) \overset{\text{def}}{=}\left(
H\circ \Lambda ^{\text{bit}}\circ H^{\dagger }\right) \left( \rho \right)
=\left( 1-p\right) \rho +pZ\rho Z\text{,}
\end{equation}%
where,%
\begin{equation}
\Lambda ^{\text{bit}}\left( \rho \right) \overset{\text{def}}{=}\left(
1-p\right) \rho +pX\rho X\text{ and, }H\overset{\text{def}}{=}\frac{1}{\sqrt{%
2}}\left( 
\begin{array}{cc}
1 & 1 \\ 
1 & -1%
\end{array}%
\right) \text{.}
\end{equation}%
These operations may be trivially incorporated into the encoding and
error-correction operations. The encoding for the phase flip channel is
performed in two steps: i) first, we encode in three qubits exactly as for
the bit flip channel; ii) second, we apply a Hadamard gate to each qubit,%
\begin{equation}
\left\vert 0\right\rangle \overset{\text{tensoring}}{\longrightarrow }%
\left\vert 000\right\rangle \overset{U_{\text{enc}}^{\text{bit}}}{%
\longrightarrow }\left\vert 000\right\rangle \overset{H^{\otimes 3}}{%
\longrightarrow }\left\vert 0_{\text{L}}\right\rangle \overset{\text{def}}{=}%
\left\vert +++\right\rangle \text{, }\left\vert 1\right\rangle \overset{%
\text{tensoring}}{\longrightarrow }\left\vert 100\right\rangle \overset{U_{%
\text{enc}}^{\text{bit}}}{\longrightarrow }\left\vert 111\right\rangle 
\overset{H^{\otimes 3}}{\longrightarrow }\left\vert 1_{\text{L}%
}\right\rangle \overset{\text{def}}{=}\left\vert ---\right\rangle \text{,}
\end{equation}%
where,%
\begin{equation}
\left( 
\begin{array}{c}
\left\vert +\right\rangle \\ 
\left\vert -\right\rangle%
\end{array}%
\right) =\frac{1}{\sqrt{2}}\left( 
\begin{array}{cc}
1 & 1 \\ 
1 & -1%
\end{array}%
\right) \left( 
\begin{array}{c}
\left\vert 0\right\rangle \\ 
\left\vert 1\right\rangle%
\end{array}%
\right) \text{.}  \label{+}
\end{equation}%
The unitary encoding operator $U_{\text{enc}}^{\text{phase}}$ is given by $%
U_{\text{enc}}^{\text{phase}}\overset{\text{def}}{=}H^{\otimes 3}\circ U_{%
\text{enc}}^{\text{bit}}$,with $U_{\text{enc}}^{\text{bit}}$ defined in (\ref%
{bitencoding}). Furthermore, in the phase flip code, the recovery operation
is the Hadamard conjugated recovery operation from the bit flip code, $%
R_{k}^{\text{phase}}\overset{\text{def}}{=}H^{\otimes 3}R_{k}^{\text{bit}%
}H^{\otimes 3}$.

\subsubsection{Model II}

We consider $\Lambda _{\mu }^{\left( n\right) }\left( \rho \right) $ in (\ref%
{gigi}) with $n=3$. Technical details will be omitted. They can be obtained
by following the line of reasoning presented for the three-qubit bit flip
error correction scheme applied to the model I. It turns out that,%
\begin{equation}
\mathcal{F}_{\text{bit}}^{\left( 3\right) }\left( \mu \text{, }p\right) =\mu
\left( -3p^{3}+6p^{2}-3p\right) +\left( 2p^{3}-3p^{2}+1\right) \text{ (model
II).}  \label{bit22}
\end{equation}

\subsection{DFS for bit flip noise}

\subsubsection{Model I}

\emph{Error Operators}. Consider the limiting case of (\ref{n-general}) with 
$n=2$ qubits and correlated errors in a bit flip quantum channel,%
\begin{equation}
\Lambda ^{(2)}(\rho )\overset{\text{def}}{=}\sum_{i_{1}\text{, }%
i_{2}=0}^{1}p_{i_{2}|i_{1}}p_{i_{1}}\left( A_{i_{2}}\otimes A_{i_{1}}\right)
\rho \left( A_{i_{2}}\otimes A_{i_{1}}\right) ^{\dagger }\text{,}
\label{bit2}
\end{equation}%
The error superoperator $\mathcal{A}$ associated to channel (\ref{bit2}) is
defined in terms of the following error operators,%
\begin{equation}
\mathcal{A}\longleftrightarrow \left\{ A_{0}^{\prime }\text{,.., }%
A_{3}^{\prime }\right\} \text{ with }\Lambda ^{(2)}(\rho )\overset{\text{def}%
}{=}\sum\limits_{k=0}^{3}A_{k}^{\prime }\rho A_{k}^{\prime \dagger }\text{
and, }\sum\limits_{k=0}^{3}A_{k}^{\prime \dagger }A_{k}^{\prime
}=I_{4\times 4}\text{.}  \label{not}
\end{equation}%
In an explicit way, the error operators $\left\{ A_{0}^{\prime }\text{,.., }%
A_{3}^{\prime }\right\} $ are given by,%
\begin{equation}
A_{0}^{\prime }=\sqrt{\tilde{p}_{0}^{\left( 2\right) }}I^{1}\otimes I^{2}%
\text{, }A_{1}^{\prime }=\sqrt{\tilde{p}_{1}^{\left( 2\right) }}X^{1}\otimes
I^{2}\text{, }A_{2}^{\prime }=\sqrt{\tilde{p}_{2}^{\left( 2\right) }}%
I^{1}\otimes X^{2}\text{, }A_{3}^{\prime }=\sqrt{\tilde{p}_{3}^{\left(
2\right) }}X^{1}\otimes X^{2}\text{, }  \label{1a}
\end{equation}%
where the coefficients $\tilde{p}_{k}^{\left( 2\right) }$ for $k=0$,.., $3$
are given by,%
\begin{equation}
\tilde{p}_{0}^{\left( 2\right) }=p_{00}p_{0}\text{, }\tilde{p}_{1}^{\left(
2\right) }=p_{10}p_{0}\text{, }\tilde{p}_{2}^{\left( 2\right) }=p_{01}p_{1}%
\text{, }\tilde{p}_{3}^{\left( 2\right) }=p_{11}p_{1}\text{.}  \label{caz}
\end{equation}

\emph{Encoding Operator}. We encode our logical qubit with a simple
decoherence free subspace of two qubits given by \cite{lidar03},%
\begin{equation}
\left\vert 0\right\rangle \longrightarrow \left\vert 0_{L}\right\rangle
=\left\vert +-\right\rangle \text{ and, }\left\vert 1\right\rangle
\longrightarrow \left\vert 1_{L}\right\rangle =\left\vert -+\right\rangle 
\text{,}  \label{pac}
\end{equation}%
with $\left\vert \pm \right\rangle $ defined in (\ref{+}). As a side remark,
we point out that a suitable DFS encoding for a phase flip Markovian
correlated noise model is given by $\left\vert 0\right\rangle
\longrightarrow \left\vert 0_{L}\right\rangle =\left\vert 01\right\rangle $
and, $\left\vert 1\right\rangle \longrightarrow \left\vert
1_{L}\right\rangle =\left\vert 10\right\rangle $.

\emph{Correctable Errors and} \emph{Recovery Operators}. The set of error
operators satisfying the detectability condition \cite{knill02}, $P_{%
\mathcal{C}}A_{k}^{\prime }P_{\mathcal{C}}=\lambda _{A_{k}^{\prime }}P_{%
\mathcal{C}}$, where $P_{\mathcal{C}}=\left\vert 0_{L}\right\rangle
\left\langle 0_{L}\right\vert +$ $\left\vert 1_{L}\right\rangle \left\langle
1_{L}\right\vert $ is the projector operator on the code subspace $\mathcal{C%
}=Span\left\{ \left\vert 0_{L}\right\rangle \text{, }\left\vert
1_{L}\right\rangle \right\} $ is given by,%
\begin{equation}
\mathcal{A}_{\text{detectable}}=\left\{ A_{0}^{\prime }\text{, }%
A_{3}^{\prime }\right\} \subseteq \mathcal{A}\text{.}
\end{equation}%
Furthermore, since all the detectable errors are invertible, the set of
correctable errors is such that $\mathcal{A}_{\text{correctable}}^{\dagger }%
\mathcal{A}_{\text{correctable}}$ is detectable. It follows then that,%
\begin{equation}
\mathcal{A}_{\text{correctable}}=\mathcal{A}_{\text{detectable}}\subseteq 
\mathcal{A}\text{.}
\end{equation}%
The action of the correctable error operators $\mathcal{A}_{\text{correctable%
}}$ on the codewords $\left\vert 0_{L}\right\rangle $ and $\left\vert
1_{L}\right\rangle $ is given by,%
\begin{equation}
\left\vert 0_{L}\right\rangle \rightarrow A_{0}^{\prime }\left\vert
0_{L}\right\rangle =\sqrt{\tilde{p}_{0}^{\left( 2\right) }}\left\vert
+-\right\rangle \text{, }A_{3}^{\prime }\left\vert 0_{L}\right\rangle =-%
\sqrt{\tilde{p}_{3}^{\left( 2\right) }}\left\vert +-\right\rangle \text{, }%
\left\vert 1_{L}\right\rangle \rightarrow A_{0}^{\prime }\left\vert
1_{L}\right\rangle =\sqrt{\tilde{p}_{0}^{\left( 2\right) }}\left\vert
-+\right\rangle \text{, }A_{3}^{\prime }\left\vert 1_{L}\right\rangle =-%
\sqrt{\tilde{p}_{3}^{\left( 2\right) }}\left\vert -+\right\rangle \text{.}
\label{ea1}
\end{equation}%
The two \ one-dimensional orthogonal subspaces $\mathcal{V}^{0_{L}}$ and $%
\mathcal{V}^{1_{L}}$ of $\mathcal{H}_{2}^{2}$ generated by the action of $%
\mathcal{A}_{\text{correctable}}$ on $\left\vert 0_{L}\right\rangle $ and $%
\left\vert 1_{L}\right\rangle $ are given by,%
\begin{equation}
\mathcal{V}^{0_{L}}=Span\left\{ \left\vert v_{1}^{0_{L}}\right\rangle
=\left\vert +-\right\rangle \right\} \text{ and, }\mathcal{V}%
^{1_{L}}=Span\left\{ \left\vert v_{1}^{1_{L}}\right\rangle =\left\vert
-+\right\rangle \right\} \text{. }
\end{equation}
Notice that $\mathcal{V}^{0_{L}}\oplus \mathcal{V}^{1_{L}}\neq \mathcal{H}%
_{2}^{2}$. This means that the trace preserving recovery superoperator $%
\mathcal{R}$ is defined in terms of one standard recovery operator $R_{1}$
and by the projector $R_{\perp }$ onto the orthogonal complement of $%
\bigoplus\limits_{i=0}^{1}\ \mathcal{V}^{i_{L}}$, i. e. the part of the
Hilbert space $\mathcal{H}_{2}^{2}$ which is not reached by acting on the
code $\mathcal{C}\ $\ with the correctable error operators. In the case
under consideration,%
\begin{equation}
R_{1}\overset{\text{def}}{=}\left\vert +-\right\rangle \left\langle
+-\right\vert +\left\vert -+\right\rangle \left\langle -+\right\vert \text{, 
}R_{\perp }=\sum_{s=1}^{2}\left\vert r_{s}\right\rangle \left\langle
r_{s}\right\vert \text{,}  \label{dfsr}
\end{equation}%
where $\left\{ \left\vert r_{s}\right\rangle \right\} $ is an orthonormal
basis for $\left( \mathcal{V}^{0_{L}}\oplus \mathcal{V}^{1_{L}}\right)
^{\perp }$. A suitable basis $\mathcal{B}_{\left( \mathcal{V}^{0_{L}}\oplus 
\mathcal{V}^{1_{L}}\right) ^{\perp }}$ is given by,%
\begin{equation}
\mathcal{B}_{\left( \mathcal{V}^{0_{L}}\oplus \mathcal{V}^{1_{L}}\right)
^{\perp }}=\left\{ r_{1}=\left\vert ++\right\rangle \text{, }%
r_{2}=\left\vert --\right\rangle \right\} \text{.}
\end{equation}%
Therefore, $\mathcal{R}\leftrightarrow \left\{ R_{1}\text{, }R_{\perp
}\right\} $ is indeed a trace preserving quantum operation,%
\begin{equation}
R_{1}^{\dagger }R_{1}+R_{\perp }^{\dagger }R_{\perp }=I_{4\times 4}\text{.}
\end{equation}%
The action of this recovery operation $\mathcal{R}$ with $R_{2}\equiv
R_{\perp }$ on the map $\Lambda ^{\left( 2\right) }\left( \rho \right) $ in (%
\ref{not}) yields,%
\begin{equation}
\Lambda _{\text{recover}}^{\left( 2\right) }\left( \rho \right) \equiv
\left( \mathcal{R\circ }\Lambda ^{(2)}\right) \left( \rho \right) \overset{%
\text{def}}{=}\sum_{k=0}^{3}\sum\limits_{l=1}^{2}\left( R_{l}A_{k}^{\prime
}\right) \rho \left( R_{l}A_{k}^{\prime }\right) ^{\dagger }\text{,}
\label{pla4}
\end{equation}

\emph{Entanglement Fidelity}. We want to describe the action of $\mathcal{%
R\circ }\Lambda ^{(2)}$ restricted to the code subspace $\mathcal{C}$.
Therefore, we compute the $2\times 2$ matrix representation $\left[
R_{l}A_{k}^{\prime }\right] _{|\mathcal{C}}$ of each $R_{l}A_{k}^{\prime }$
with $l=1$, $2$ and $k=0$,.., $3$ where,%
\begin{equation}
\left[ R_{l}A_{k}^{\prime }\right] _{|\mathcal{C}}\overset{\text{def}}{=}%
\left( 
\begin{array}{cc}
\left\langle 0_{L}|R_{l}A_{k}^{\prime }|0_{L}\right\rangle  & \left\langle
0_{L}|R_{l}A_{k}^{\prime }|1_{L}\right\rangle  \\ 
\left\langle 1_{L}|R_{l}A_{k}^{\prime }|0_{L}\right\rangle  & \left\langle
1_{L}|R_{l}A_{k}^{\prime }|1_{L}\right\rangle 
\end{array}%
\right) \text{.}  \label{sup1}
\end{equation}

\begin{figure}[htpb]
\begin{center}
\includegraphics[scale=0.9]{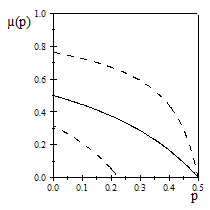}
\end{center}
\vskip-12pt
\caption{Threhold curves for model I:
concatenated code (dashed line), DFS (thin solid line).}
\label{Figure-1}
\end{figure}

Substituting (\ref{1a}) and (\ref%
{dfsr}) into (\ref{sup1}), it turns out that the only matrices $\left[
R_{l}A_{k}^{\prime }\right] _{|\mathcal{C}}$ with non-vanishing trace are
given by,%
\begin{equation}
\left[ R_{1}A_{0}^{\prime }\right] _{|\mathcal{C}}=\sqrt{\tilde{p}%
_{0}^{\left( 2\right) }}\left( 
\begin{array}{cc}
1 & 0 \\ 
0 & 1%
\end{array}%
\right) \text{, }\left[ R_{1}A_{3}^{\prime }\right] _{|\mathcal{C}}=-\sqrt{%
\tilde{p}_{3}^{\left( 2\right) }}\left( 
\begin{array}{cc}
1 & 0 \\ 
0 & 1%
\end{array}%
\right) 
\end{equation}%
Therefore, the entanglement fidelity $\mathcal{F}_{DFS}^{\left( 2\right)
}\left( \mu \text{, }p\right) $ defined as,%
\begin{equation}
\mathcal{F}_{DFS}^{\left( 2\right) }\left( \mu \text{, }p\right) \overset{%
\text{def}}{=}\mathcal{F}^{\left( 2\right) }\left( \frac{1}{2}I_{2\times 2}%
\text{, }\mathcal{R\circ }\Lambda ^{(2)}\right) =\frac{1}{\left( 2\right)
^{2}}\sum_{k=0}^{3}\sum\limits_{l=1}^{2}\left\vert \text{tr}\left( \left[
R_{l}A_{k}^{\prime }\right] _{|\mathcal{C}}\right) \right\vert ^{2}\text{,}
\label{pla3}
\end{equation}%
results,%
\begin{equation}
\mathcal{F}_{DFS}^{\left( 2\right) }\left( \mu \text{, }p\right) =\tilde{p}%
_{0}^{\left( 2\right) }+\tilde{p}_{3}^{\left( 2\right) }\text{.}
\label{dfsusa}
\end{equation}%
The expression for $\mathcal{F}_{DFS}^{\left( 2\right) }\left( \mu \text{, }%
p\right) $ in (\ref{pla3}) represents the entanglement fidelity quantifying
the performance of the error correction scheme provided by the error
avoiding code here considered. The quantum operation $\mathcal{R\circ }%
\Lambda ^{(2)}$ appearing in (\ref{pla3}) is defined in equation (\ref{pla4}%
) and the recovery operators $R_{l}$ are explicitly given in (\ref{dfsr}).
The action of\textbf{\ }$R_{l}A_{k}^{\prime }$ in (\ref{pla3}) is restricted
to the code space\textbf{\ }$\mathcal{C}$ defined in (\ref{pac}).

Substituting (\ref{usa2}) and (\ref{caz}) into (\ref{dfsusa}), we finally
obtain%
\begin{equation}
\mathcal{F}_{DFS}^{\left( 2\right) }\left( \mu \text{, }p\right) =\mu \left(
-2p^{2}+2p\right) +\left( 2p^{2}-2p+1\right) \text{ (model I).}  \label{ef1}
\end{equation}%
We point out that error correction schemes improve the transmission accuracy
only if the failure probability $\mathcal{P}\left( \mu \text{, }p\right) $
is strictly less than the error probability $p$ \cite{gaitan},

\begin{equation}
\mathcal{P}\left( \mu \text{, }p\right) \overset{\text{def}}{=}1-\mathcal{F}%
\left( \mu \text{, }p\right) <p\text{.}  \label{curve}
\end{equation}%
In view of (\ref{curve}), we can determine threshold curves $\bar{\mu}\left(
p\right) $ that allow to select the two-dimensional parametric region where
error correction schemes are useful. For instance, considering the model I
in absence of correlations, it follows that the three-qubit code is
effective only if $p<0.5$. However for such values of the error probability,
the DFS considered does not work for $\mu $ approaching zero. The threshold
curve for the DFS for the model I appear in Figure $1$. The DFS works only
in the parametric region above the thin solid line in Figure 1, while the
three-qubit bit-flip code works for all values of the memory degree $\mu $
when the error probability is less than $0.5$.

\subsubsection{Model II}

We consider $\Lambda _{\mu }^{\left( n\right) }\left( \rho \right) $ in (\ref%
{gigi}) with $n=2$. Technical details will be omitted. They can be obtained
by following the line of reasoning presented when studying the error
avoiding code applied to the model I. It turns out that,%
\begin{equation}
\mathcal{F}_{DFS}^{\left( 2\right) }\left( \mu \text{, }p\right) =\mu \left(
-2p^{2}+2p\right) +\left( 2p^{2}-2p+1\right) \text{ (model II).}  \label{ef2}
\end{equation}%
Notice that the entanglement fidelities in (\ref{ef1}) and (\ref{ef2}) are
equal. Finally, the threshold curves for the three-qubit bit flip and DFS
for the model II appear in Figure $2$. The DFS works only in the parametric
region above the dashed line while the three-qubit code works only below the
thin solid line. 

\begin{figure}[htpb]
\begin{center}
\includegraphics[scale=0.9]{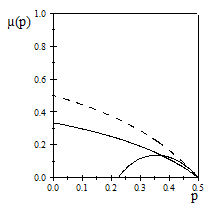}
\end{center}
\vskip-12pt
\caption{Threshold curves
for model II: DFS (dashed line), three-qubit bit flip code (thin solid
line), concatenated code (thick solid line). }
\label{Figure-2}
\end{figure}

\section{Concatenated Codes}

\subsection{Model I}

\emph{Error Operators}. Consider the limiting case of (\ref{n-general}) with 
$n=6$ qubits and correlated errors in a bit flip quantum channel,%
\begin{equation}
\Lambda ^{(6)}(\rho )\overset{\text{def}}{=}\sum_{i_{1}\text{,..., }%
i_{6}=0}^{1}p_{i_{6}|i_{5}}p_{i_{5}|i_{4}}p_{i_{4}|i_{3}}p_{i_{3}|i_{2}}p_{i_{2}|i_{1}}p_{i_{1}}\left( A_{i_{6}}A_{i_{5}}A_{i_{4}}A_{i_{3}}A_{i_{2}}A_{i_{1}}\right) \rho \left( A_{i_{6}}A_{i_{5}}A_{i_{4}}A_{i_{3}}A_{i_{2}}A_{i_{1}}\right) ^{\dagger }%
\text{,}  \label{rome}
\end{equation}%
The error superoperator $\mathcal{A}$ associated to channel (\ref{rome}) is
defined in terms of the following error operators,%
\begin{equation}
\mathcal{A}\longleftrightarrow \left\{ A_{0}^{\prime }\text{,.., }%
A_{63}^{\prime }\right\} \text{ with }\Lambda ^{(6)}(\rho )\overset{\text{def%
}}{=}\sum\limits_{k=0}^{2^{6}-1}A_{k}^{\prime }\rho A_{k}^{\prime \dagger }%
\text{ and, }\sum\limits_{k=0}^{2^{6}-1}A_{k}^{\prime \dagger
}A_{k}^{\prime }=I_{64\times 64}\text{.}  \label{KR}
\end{equation}%
The error operators in the Kraus decomposition (\ref{KR}) are $2^{6}=64$,%
\begin{equation}
\sum_{k=0}^{6}\binom{6}{k}=2^{6}\text{,}
\end{equation}%
where $\binom{6}{k}$ is the cardinality of weight-$k$ error operators.

\emph{Encoding Operator}. We encode our logical qubit with a concatenated
subspace obtained by combining the decoherence free subspace in (\ref{pac})
(inner code, $\mathcal{C}_{DFS}=\mathcal{C}_{\text{inner}}$ ) with the
three-qubit bit flip code in (\ref{placs}) (outer code, $\mathcal{C}_{\text{%
bit}}=\mathcal{C}_{\text{outer}}$). We obtain that the codewords of the
concatenated code $\mathcal{C}=\mathcal{C}_{DFS}\circ \mathcal{C}_{\text{bit}%
}$ are given by,%
\begin{eqnarray}
\left\vert 0_{L}\right\rangle &=&\frac{1}{2}\left( \left\vert
000000\right\rangle -\left\vert 000111\right\rangle +\left\vert
111000\right\rangle -\left\vert 111111\right\rangle \right) \text{,}  \notag
\\
&&  \notag \\
\left\vert 1_{L}\right\rangle &=&\frac{1}{2}\left( \left\vert
000000\right\rangle +\left\vert 000111\right\rangle -\left\vert
111000\right\rangle -\left\vert 111111\right\rangle \right) \text{.}
\label{CW}
\end{eqnarray}%
As a side remark, we point out that a suitable concatenated code for the
case of a phase flip Markovian correlated noise model is given by $%
\left\vert 0\right\rangle \longrightarrow \left\vert 0_{L}\right\rangle
=\left\vert +++---\right\rangle $ and, $\left\vert 1\right\rangle
\longrightarrow \left\vert 1_{L}\right\rangle =\left\vert
---+++\right\rangle $.

\emph{Correctable Errors and Recovery Operators}. Recall that the
detectability condition is given by $P_{\mathcal{C}}A_{k}^{\prime }P_{%
\mathcal{C}}=\lambda _{A_{k}^{\prime }}P_{\mathcal{C}}$ where the projector
operator on the code space $\mathcal{C}$ is $P_{\mathcal{C}}=\left\vert
0_{L}\right\rangle \left\langle 0_{L}\right\vert +\left\vert
1_{L}\right\rangle \left\langle 1_{L}\right\vert $. Observe that,%
\begin{equation}
P_{\mathcal{C}}A_{k}^{\prime }P_{\mathcal{C}}=\left\langle
0_{L}|A_{k}^{\prime }|0_{L}\right\rangle \left\vert 0_{L}\right\rangle
\left\langle 0_{L}\right\vert +\left\langle 0_{L}|A_{k}^{\prime
}|1_{L}\right\rangle \left\vert 0_{L}\right\rangle \left\langle
1_{L}\right\vert +\left\langle 1_{L}|A_{k}^{\prime }|0_{L}\right\rangle
\left\vert 1_{L}\right\rangle \left\langle 0_{L}\right\vert +\left\langle
1_{L}|A_{k}^{\prime }|1_{L}\right\rangle \left\vert 1_{L}\right\rangle
\left\langle 1_{L}\right\vert \text{.}
\end{equation}%
Therefore, it turns out that for detectable error operators we must have,%
\begin{equation}
\left\langle 0_{L}|A_{k}^{\prime }|0_{L}\right\rangle =\left\langle
1_{L}|A_{k}^{\prime }|1_{L}\right\rangle \text{ and, }\left\langle
0_{L}|A_{k}^{\prime }|1_{L}\right\rangle =\left\langle 1_{L}|A_{k}^{\prime
}|0_{L}\right\rangle =0\text{.}
\end{equation}%
In the case under consideration, it follows that the only error operators
(omitting for the sake of simplicity the proper error amplitudes) not
fulfilling the above conditions are proportional to,%
\begin{equation}
X^{1}X^{2}X^{3}\text{ and }X^{4}X^{5}X^{6}\text{.}
\end{equation}%
For such operators, we get%
\begin{equation}
\left\langle 0_{L}|X^{1}X^{2}X^{3}|0_{L}\right\rangle =1\text{, }%
\left\langle 1_{L}|X^{1}X^{2}X^{3}|1_{L}\right\rangle =-1\text{, and, }%
\left\langle 0_{L}|X^{4}X^{5}X^{6}|0_{L}\right\rangle =-1\text{, }%
\left\langle 1_{L}|X^{4}X^{5}X^{6}|1_{L}\right\rangle =1\text{.}
\end{equation}%
Therefore $X^{1}X^{2}X^{3}$ and $X^{4}X^{5}X^{6}$ are not detectable. Thus,
the cardinality of the set of detectable errors $\mathcal{A}_{\text{%
detectable}}$ is $62$. Furthermore, recall that the set of correctable
errors $\mathcal{A}_{\text{correctable}}$ is such that $\mathcal{A}_{\text{%
correctable}}^{\dagger }\mathcal{A}_{\text{correctable}}$ is detectable (in
the hypothesis of invertible error operators). Therefore, after some
reasoning, we conclude that the set of correctable errors is composed by $32$
error operators. The correctable weight-$0$, $1$ and $2$ correctable error
operators are (omitting the proper error amplitudes),%
\begin{equation}
\left\{ I\right\} _{\text{weight-}0}\text{, }\left\{ X^{1}\text{, }X^{2}%
\text{, }X^{3}\text{, }X^{4}\text{, }X^{5}\text{, }X^{6}\right\} _{\text{%
weight-}1}\text{, }
\end{equation}%
and,%
\begin{equation}
\left\{ X^{1}X^{4}\text{, }X^{1}X^{5}\text{, }X^{1}X^{6}\text{, }X^{2}X^{4}%
\text{, }X^{2}X^{5}\text{, }X^{2}X^{6}\text{, }X^{3}X^{4}\text{, }X^{3}X^{5}%
\text{, }X^{3}X^{6}\right\} _{\text{weight-}2}\text{,}
\end{equation}%
respectively. The correctable weight-$4$ errors are,%
\begin{equation}
\left\{ 
\begin{array}{c}
X^{1}X^{2}X^{4}X^{5}\text{, }X^{1}X^{2}X^{4}X^{6}\text{, }%
X^{1}X^{2}X^{5}X^{6}\text{, }X^{1}X^{3}X^{4}X^{5}\text{, }%
X^{1}X^{3}X^{4}X^{6}\text{, } \\ 
\\ 
X^{1}X^{3}X^{5}X^{6}\text{, }X^{2}X^{3}X^{4}X^{5}\text{, }%
X^{2}X^{3}X^{4}X^{6}\text{, }X^{2}X^{3}X^{5}X^{6}%
\end{array}%
\right\} _{\text{weight-}4}\text{.}
\end{equation}%
Finally, weight-$5$ and weight-$6$ error operators are given by, 
\begin{equation}
\left\{ X^{1}X^{2}X^{3}X^{4}X^{5}\text{, }X^{1}X^{2}X^{3}X^{4}X^{6}\text{, }%
X^{1}X^{2}X^{4}X^{5}X^{6}\text{, }X^{1}X^{3}X^{4}X^{5}X^{6}\text{, }%
X^{1}X^{2}X^{3}X^{5}X^{6}\text{, }X^{2}X^{3}X^{4}X^{5}X^{6}\text{ }\right\}
_{\text{weight-}5}\text{,}
\end{equation}%
and,%
\begin{equation}
\left\{ X^{1}X^{2}X^{3}X^{4}X^{5}X^{6}\right\} _{\text{weight-}6}\text{,}
\end{equation}%
respectively. The action of the correctable errors on the codewords in (\ref%
{CW}) is such that the Hilbert space $\mathcal{H}_{2}^{6}$ can be decomposed
in two $32$-dimensional orthogonal subspaces $\mathcal{V}^{0_{L}}$ and $%
\mathcal{V}^{1_{L}}$. In other words, $\mathcal{H}_{2}^{6}=\mathcal{V}%
^{0_{L}}\oplus \mathcal{V}^{1_{L}}$ where%
\begin{equation}
\mathcal{V}^{0_{L}}=Span\left\{ \left\vert v_{k+1}^{0_{L}}\right\rangle =%
\frac{1}{\sqrt{\tilde{p}_{k}^{\left( 6\right) }}}A_{k}^{\prime }\left\vert
0_{L}\right\rangle \right\} \text{ and, }\mathcal{V}^{1_{L}}=Span\left\{
\left\vert v_{k+1}^{1_{L}}\right\rangle =\frac{1}{\sqrt{\tilde{p}%
_{k}^{\left( 6\right) }}}A_{k}^{\prime }\left\vert 1_{L}\right\rangle
\right\} \text{,}
\end{equation}%
with $A_{k}^{\prime }\in \mathcal{A}_{\text{correctable}}$ $\forall k=0$%
,..., $31$ (numbering the correctable error operators from $0$ to $31$).
Notice that $\left\langle v_{k}^{i_{L}}|v_{k^{\prime }}^{j_{L}}\right\rangle
=\delta _{kk^{\prime }}\delta _{ij}$, with $k$, $k^{\prime }\in \left\{ 0%
\text{,..., }31\right\} $ and $i$, $j\in \left\{ 0\text{, }1\right\} $ since,%
\begin{equation}
\left\langle v_{k}^{i_{L}}|v_{k^{\prime }}^{j_{L}}\right\rangle
=\left\langle i_{L}|\frac{A_{k-1}^{\prime \dagger }}{\sqrt{\tilde{p}%
_{k}^{\left( 6\right) }}}\frac{A_{k^{\prime }-1}^{\prime }}{\sqrt{\tilde{p}%
_{k^{\prime }}^{\left( 6\right) }}}|j_{L}\right\rangle =\frac{1}{\sqrt{%
\tilde{p}_{k}^{\left( 6\right) }\tilde{p}_{k^{\prime }}^{\left( 6\right) }}}%
\left\langle i_{L}|A_{k}^{\prime \dagger }A_{k^{\prime }}^{\prime
}|j_{L}\right\rangle =\frac{1}{\sqrt{\tilde{p}_{k}^{\left( 6\right) }\tilde{p%
}_{k^{\prime }}^{\left( 6\right) }}}\alpha _{kk^{\prime }}^{\prime }\delta
_{ij}=\delta _{kk^{\prime }}\delta _{ij}\text{,}
\end{equation}%
where we have used the fact that the square (Hermitian) matrix $\alpha
_{kk^{\prime }}^{\prime }$ equals $\sqrt{\tilde{p}_{k}^{\left( 6\right) }%
\tilde{p}_{k^{\prime }}^{\left( 6\right) }}$ $\delta _{kk^{\prime }}$. The
recovery superoperator $\mathcal{R}\leftrightarrow \left\{ R_{l}\right\} $
with $l=1$,.., $32$ is defined as \cite{knill97},%
\begin{equation}
R_{l}\overset{\text{def}}{=}V_{l}\sum_{i=0}^{1}\left\vert
v_{l}^{i_{L}}\right\rangle \left\langle v_{l}^{i_{L}}\right\vert \text{,}
\end{equation}%
where the unitary operator $V_{l}$ is such that $V_{l}\left\vert
v_{l}^{i_{L}}\right\rangle =\left\vert i_{L}\right\rangle $ for $i\in
\left\{ 0\text{, }1\right\} $. Notice that,%
\begin{equation}
R_{l}\overset{\text{def}}{=}V_{l}\sum_{i=0}^{1}\left\vert
v_{l}^{i_{L}}\right\rangle \left\langle v_{l}^{i_{L}}\right\vert =\left\vert
0_{L}\right\rangle \left\langle v_{l}^{0_{L}}\right\vert +\left\vert
1_{L}\right\rangle \left\langle v_{l}^{1_{L}}\right\vert \text{.}
\end{equation}%
If turns out that the $32$ recovery operators are given by,%
\begin{equation}
R_{l+1}=R_{1}\frac{A_{l}^{\prime }}{\sqrt{\tilde{p}_{l}^{\prime }}}=\left(
\left\vert 0_{L}\right\rangle \left\langle 0_{L}\right\vert +\left\vert
1_{L}\right\rangle \left\langle 1_{L}\right\vert \right) \frac{A_{l}^{\prime
}}{\sqrt{\tilde{p}_{l}^{\prime }}}\text{,}
\end{equation}%
with $l\in \left\{ 0\text{,..., }31\right\} $. Notice that $\mathcal{R}%
\leftrightarrow \left\{ R_{l}\right\} $ is a trace preserving quantum
operation since,%
\begin{equation}
\sum_{l=1}^{32}R_{l}^{\dagger }R_{l}=\sum_{l=1}^{32}\left(
\sum_{i_{L}}\left\vert i_{L}\right\rangle \left\langle
v_{l}^{i_{L}}\right\vert \right) ^{\dagger }\left( \sum_{j_{L}}\left\vert
j_{L}\right\rangle \left\langle v_{l}^{j_{L}}\right\vert \right)
=\sum_{l=1}^{32}\sum_{i_{L\text{, }}j_{L}}\left\vert
v_{l}^{i_{L}}\right\rangle \left\langle i_{L}|j_{L}\right\rangle
\left\langle v_{l}^{j_{L}}\right\vert =\sum_{l=1}^{32}\sum_{i_{L}}\left\vert
v_{l}^{i_{L}}\right\rangle \left\langle v_{l}^{i_{L}}\right\vert
=I_{64\times 64}\text{,}
\end{equation}%
since $\left\{ \left\vert v_{l}^{i_{L}}\right\rangle \right\} $ with $l=1$%
,..., $32$ and $i_{L}\in \left\{ 0\text{, }1\right\} $ is an orthonormal
basis for $\mathcal{H}_{2}^{6}$. Finally, the action of this recovery
operation $\mathcal{R}$ on the map $\Lambda ^{\left( 6\right) }\left( \rho
\right) $ in (\ref{KR}) leads to,%
\begin{equation}
\Lambda _{\text{recover}}^{\left( 6\right) }\left( \rho \right) \equiv
\left( \mathcal{R\circ }\Lambda ^{(6)}\right) \left( \rho \right) \overset{%
\text{def}}{=}\sum_{k=0}^{2^{6}-1}\sum\limits_{l=1}^{32}\left(
R_{l}A_{k}^{\prime }\right) \rho \left( R_{l}A_{k}^{\prime }\right)
^{\dagger }\text{.}
\end{equation}

\emph{Entanglement Fidelity}. We want to describe the action of $\mathcal{%
R\circ }\Lambda ^{(6)}$ restricted to the code subspace $\mathcal{C}$.
Recalling that $A_{l}^{\prime }=A_{l}^{\prime \dagger }$, it turns out that,%
\begin{equation}
\left\langle i_{L}|R_{l+1}A_{k}^{\prime }|j_{L}\right\rangle =\frac{1}{\sqrt{%
\tilde{p}_{l}^{\prime }}}\left\langle i_{L}|0_{L}\right\rangle \left\langle
0_{L}|A_{l}^{\prime \dagger }A_{k}^{\prime }|j_{L}\right\rangle +\frac{1}{%
\sqrt{\tilde{p}_{l}^{\prime }}}\left\langle i_{L}|1_{L}\right\rangle
\left\langle 1_{L}|A_{l}^{\prime \dagger }A_{k}^{\prime }|j_{L}\right\rangle 
\text{.}
\end{equation}%
We now need to compute the $2\times 2$ matrix representation $\left[
R_{l}A_{k}^{\prime }\right] _{|\mathcal{C}}$ of each $R_{l}A_{k}^{\prime }$
with $l=0$,.., $31$ and $k=0$,.., $2^{6}-1$ where,%
\begin{equation}
\left[ R_{l+1}A_{k}^{\prime }\right] _{|\mathcal{C}}\overset{\text{def}}{=}%
\left( 
\begin{array}{cc}
\left\langle 0_{L}|R_{l+1}A_{k}^{\prime }|0_{L}\right\rangle  & \left\langle
0_{L}|R_{l+1}A_{k}^{\prime }|1_{L}\right\rangle  \\ 
\left\langle 1_{L}|R_{l+1}A_{k}^{\prime }|0_{L}\right\rangle  & \left\langle
1_{L}|R_{l+1}A_{k}^{\prime }|1_{L}\right\rangle 
\end{array}%
\right) \text{.}
\end{equation}%
For $l$, $k=0$,.., $31$, we note that $\left[ R_{l+1}A_{k}^{\prime }\right]
_{|\mathcal{C}}$ becomes,%
\begin{equation}
\left[ R_{l+1}A_{k}^{\prime }\right] _{|\mathcal{C}}=\left( 
\begin{array}{cc}
\left\langle 0_{L}|A_{l}^{\prime \dagger }A_{k}^{\prime }|0_{L}\right\rangle 
& 0 \\ 
0 & \left\langle 1_{L}|A_{l}^{\prime \dagger }A_{k}^{\prime
}|1_{L}\right\rangle 
\end{array}%
\right) =\sqrt{\tilde{p}_{l}^{\prime }}\delta _{lk}\left( 
\begin{array}{cc}
1 & 0 \\ 
0 & 1%
\end{array}%
\right) \text{,}
\end{equation}%
while for any pair $\left( l\text{, }k\right) $ with $l$ $=0$,.., $31$ and $%
k>31$, it follows that,%
\begin{equation}
\left\langle 0_{L}|R_{l+1}A_{k}^{\prime }|0_{L}\right\rangle +\left\langle
1_{L}|R_{l+1}A_{k}^{\prime }|1_{L}\right\rangle =0\text{.}
\end{equation}%
We conclude that the only matrices $\left[ R_{l}A_{k}^{\prime }\right] _{|%
\mathcal{C}}$ with non-vanishing trace are given by $\left[
R_{l+1}A_{l}^{\prime }\right] _{|\mathcal{C}}$ with $l$ $=0$,.., $31$ where,%
\begin{equation}
\left[ R_{l+1}A_{l}^{\prime }\right] _{|\mathcal{C}}=\sqrt{\tilde{p}%
_{l}^{\prime }}\left( 
\begin{array}{cc}
1 & 0 \\ 
0 & 1%
\end{array}%
\right) \text{.}
\end{equation}%
Therefore, the entanglement fidelity $\mathcal{F}_{\text{conc}}^{\left(
6\right) }\left( \mu \text{, }p\right) $ defined as,%
\begin{equation}
\mathcal{F}_{\text{conc}}^{\left( 6\right) }\left( \mu \text{, }p\right) 
\overset{\text{def}}{=}\mathcal{F}_{\text{conc}}^{\left( 6\right) }\left( 
\frac{1}{2}I_{2\times 2}\text{, }\mathcal{R\circ }\Lambda ^{(6)}\right) =%
\frac{1}{\left( 2\right) ^{2}}\sum_{k=0}^{2^{6}-1}\sum\limits_{l=1}^{32}%
\left\vert \text{tr}\left( \left[ R_{l}A_{k}^{\prime }\right] _{|\mathcal{C}%
}\right) \right\vert ^{2}\text{,}
\end{equation}%
becomes,

\begin{figure}[htpb]
\begin{center}
\includegraphics[scale=0.9]{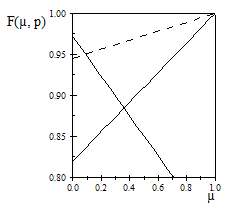}
\end{center}
\vskip-12pt
\caption{Entanglement fidelity vs.
memory parameter $\protect\mu $ with $p=10^{-1}$ for model II: concatenated
code (dashed line), DFS (thin solid line) and three-qubit bit flip code
(thick solid line).}
\label{Figure-3}
\end{figure}

\begin{eqnarray}
\mathcal{F}_{\text{conc}}^{\left( 6\right) }\left( \mu \text{, }p\right) 
&=&p_{00}^{5}p_{0}+2p_{00}^{4}p_{10}p_{0}+p_{00}^{3}p_{01}p_{10}\left(
4p_{0}+p_{1}\right)
+4p_{00}^{2}p_{01}p_{10}^{2}p_{0}+p_{00}^{2}p_{01}p_{10}p_{11}p_{0}+3p_{00}p_{01}^{2}p_{10}^{2}p_{0}+
\notag \\
&&  \notag \\
&&+4p_{01}p_{10}^{2}p_{11}^{2}p_{0}+3p_{01}^{2}p_{10}^{2}p_{11}p_{1}+p_{00}p_{01}p_{10}p_{11}^{2}p_{1}+p_{01}p_{10}p_{11}^{3}\left( p_{0}+4p_{1}\right) +2p_{10}p_{11}^{4}p_{0}+p_{11}^{5}p_{1}%
\text{.}  \label{ef}
\end{eqnarray}%
Substituting (\ref{usa2}) into (\ref{ef}), we finally get%
\begin{eqnarray}
\mathcal{F}_{\text{conc}}^{\left( 6\right) }\left( \mu \text{, }p\right) 
&=&\mu ^{5}\left( -8p^{6}+24p^{5}-24p^{4}+\allowbreak 8p^{3}\right) +\mu
^{4}\left( 40p^{6}-120p^{5}+130p^{4}\allowbreak -60p^{3}+10p^{2}\right) + 
\notag \\
&&  \notag \\
&&+\mu ^{3}\left( -80p^{6}+240p^{5}-264p^{4}+128p^{3}-26p^{2}+2p\right) + 
\notag \\
&&  \notag \\
&&+\mu ^{2}\left( 80p^{6}-\allowbreak
240p^{5}+252p^{4}-104p^{3}+10p^{2}+2p\right) +  \notag \\
&&  \notag \\
&&+\mu \left( -40\allowbreak p^{6}+120p^{5}-112p^{4}+24\allowbreak
p^{3}+12\allowbreak p^{2}-4p\right) +\left(
8p^{6}-24p^{5}+18p^{4}+4p^{3}-6p^{2}+1\right) \text{.}  \label{conc1}
\end{eqnarray}%
The threshold curves for the concatenated code defined in (\ref{CW}) for the
model I appear in Figure $1$. It turns out that the concatenated code does
not work in the region delimited by the two dashed lines. We emphasize that
in view of equations (\ref{bit1}), (\ref{ef1}) and (\ref{conc1}), it turns
out that the concatenated code outperforms the bit-flip code in regions with
very high memory parameter values and outperforms the DFS in regions with
low memory parameter values. In particular,\ the relevance of the
concatenation trick shines where the weaknesses of the inner and outer codes
(yet not concatenated) are more pronounced, that is in regions with both
high error probability and low memory parameter values. It is within this
area that we can identify the region where the concatenated code outperforms
both the inner and the outer codes. However, there is a big parametric
region characterized by intermediate values of the degree of memory \ (see
Figure I) where the concatenation trick does not work well for model I. On
the contrary, we will show that the quantum coding trick is particularly
useful for the model II in the presence of partial correlations.

\subsection{Model II}

We consider $\Lambda _{\mu }^{\left( n\right) }\left( \rho \right) $ in (\ref%
{gigi}) with $n=6$. Technical details will be omitted. They can be obtained
by following the line of reasoning presented when studying the concatenated
code applied to the model I. It turns out that,%
\begin{equation}
\mathcal{F}_{\text{conc}}^{\left( 6\right) }(\mu \text{, }p)=\mu \left(
-8p^{6}+24p^{5}-18p^{4}-4p^{3}+6p^{2}\right) +\left(
8p^{6}-24p^{5}+18p^{4}+4p^{3}-6p^{2}+1\right) \text{.}  \label{EN2}
\end{equation}%
The threshold curve for the concatenated code defined in (\ref{CW}) for the
model II appears in Figure $2$. It follows that while none of the two codes
is effective in the extreme limit when the other is, the three-qubit bit
flip (phase flip) code still works for correlated errors, whereas the error
avoiding code does not work in the absence of correlations. The concatenated
code works everywhere except below the thick solid curve in Figure $2$. Here
there is a parametric region characterized by partial correlations and
delimited by the dashed and thin solid lines where only the concatenated
code is effective. Furthermore, from (\ref{bit22}), (\ref{ef2}) and (\ref%
{EN2}), it follows that the concatenated code is especially advantageous for
the model II for partially correlated error operators. For the sake of
clarity, in Figure $3$ we plot the entanglement fidelities (\ref{ef2}) (thin
solid line), (\ref{EN2}) (dashed line) and (\ref{bit22}) (tick solid line)
for $p=10^{-1}$. For such value of the error probability, the error avoiding
code only works for $\mu \gtrsim 0.44$ (threshold value obtained from Figure 
$2$), the three-qubit bit flip code only works for $\mu \lesssim 0.30$
(threshold value obtained from Figure $2$) while the concatenated code with
entanglement fidelity given in (\ref{EN2}) is is efficient for any value of $%
\mu \in \left[ 0\text{, }1\right] $.

\section{Final Remarks}

In this article, we studied the performance of simple error correcting and
error avoiding quantum codes together with their concatenation for
correlated noise models. For model I, a bit-flip (phase-flip) noisy
Markovian memory channel, we have applied both the three-qubit bit flip
(phase flip) and a suitable error avoiding code. The performance of the
codes was quantified in terms of the entanglement fidelities in (\ref{bit1})
and (\ref{ef1}). The performance of the concatenated code applied to model I
appears in (\ref{conc1}). In Figure $1$, we have plotted the parametric
regions where error correction is effective. We have presented a similar
analysis for the model II, a memory channel defined as a memory degree
dependent linear combination of memoryless channels with Kraus
decompositions expressed solely in terms of tensor products of $X$-Pauli ($Z$%
-Pauli) operators (model II). The performance of the codes was quantified in
terms of the entanglement fidelities in (\ref{bit22}) and (\ref{ef2}). The
performance of the concatenated code applied to model II appears in (\ref%
{EN2}). In Figure $2$, we have plotted the parametric regions where error
correction schemes are effective.

Our analysis explicitly shows that while none of the two codes is effective
in the extreme limit when the other is, the three-qubit bit flip (phase
flip) code still works for correlated errors, whereas the error avoiding
code does not work in the absence of correlations. Finally, our final
finding leads to conclude that the concatenated code in (\ref{CW}) is
particularly advantageous for model II in the regime of partial correlations
(see Figure $3$).

\begin{acknowledgments}
C. C. thanks S. L'Innocente and C. Lupo for useful discussions. This work
was supported by the European Community's Seventh Framework Program under
grant agreement 213681 (CORNER Project; FP7/2007-2013).
\end{acknowledgments}

\end{document}